\documentclass[11pt,a4paper,reprint,notitlepage,superscriptaddress,aps,prl]{revtex4-2}

\usepackage{graphicx}

\usepackage{amsmath,amssymb,amsfonts}

\usepackage{amsthm}

\usepackage{bbold}
\usepackage{physics}

\usepackage[dvipsnames]{xcolor}
\usepackage{graphicx}

\usepackage{hyperref}
\hypersetup{
    colorlinks,
    citecolor=red,
    filecolor=red,
    linkcolor=red,
    urlcolor=red
}

\newcommand{\ba}{\begin{eqnarray}}
\newcommand{\ea}{\end{eqnarray}}

\newcommand{\be}{\begin{equation}}
\newcommand{\ee}{\end{equation}}

\usepackage[capitalise]{cleveref}

\usepackage{cancel}
\usepackage{xcolor}

\begin{document}

\preprint{PRL/}

\title{Emergence of Spacetime from Fluctuations}

\author{Marcus Reitz}
 \affiliation{Jagiellonian University, Astronomy and Applied Computer Science, Institute of Theoretical Physics, Łojasiewicza 11, Kraków, PL 30-348, Poland.
}
\email{marcus.reitz@uj.edu.pl}

\author{Barbara \v{S}oda}
 \affiliation{Department of Physics, University of Waterloo, Waterloo, ON N2L 3G1, Canada.
}
\affiliation{Perimeter Institute for Theoretical Physics, Waterloo, ON N2L 2Y5, Canada.}
\email{bsoda@perimeterinstitute.ca}

\author{Achim Kempf}
 \affiliation{Department of Physics, University of Waterloo, Waterloo, ON N2L 3G1, Canada.
}
\affiliation{Perimeter Institute for Theoretical Physics, Waterloo, ON N2L 2Y5, Canada.}
\affiliation{Department of Applied Mathematics, University of Waterloo, Waterloo, ON N2L 3G1, Canada.}
\affiliation{Institute for Quantum Computing, University of Waterloo, Waterloo, ON N2L 3G1, Canada.}
\email{akempf@perimeterinstitute.ca}

\date{\today}

\begin{abstract} 

We use a result of Hawking and Gilkey to define a Euclidean path integral of gravity and matter 
which
has the special property of being independent of the choice of basis in the space of fields. 
This property allows the path integral to describe also physical regimes that do not admit position bases. These physical regimes are pre-geometric in the sense that they do not admit a mathematical representation of the physical degrees of freedom in terms of fields that live on a spacetime. In regimes in which a spacetime representation does emerge, the geometric properties of the emergent spacetime, such as its dimension and volume, depend on the balance of fermionic pressure 
and bosonic and gravitational pull.  That balance depends, at any given energy scale, on the number of bosonic and fermionic species that contribute, which in turn depends on their masses. This yields an explicit mechanism by which the effective spacetime dimension can depend on the energy scale. 

\end{abstract}
\maketitle

While there are a number of promising approaches to quantum gravity, see, e.g., \cite{kiefer2007quantum,Rovelli2005, Ashtekar2021,Loll2012, Loll2020,Kiefer,Kalau, Connes, Landi,ReuterSaueressig2019,Surya_2019,BeckerBeckerSchwarz2006,AmmonErdmenger2015,Oriti2020,finster2016causal,Jacobson1995, Jacobson2019, Erik2017, carlip2014black},
the question has not been settled how
a spacetime that hosts matter fields could emerge from a pre-geometric quantum gravity regime. It has also remained a largely open question how the dimension of spacetime might depend on the energy scale. 
To investigate these questions, we here work with a path integral of Euclidean signature.  The Euclidean signature is commonly used because of its technical advantages, such as better convergence properties, see, e.g., \cite{gibbons1993Euclidean,mukhanov2007introduction}. Here, we choose the Euclidean signature because it enables the use of powerful techniques of spectral geometry, such as Weyl's asymptotic formula for characterizing a spacetime's dimension, and the ability to use the gap of the Laplacian to characterize the size of a spacetime. Our general approach does not depend on the choice of the Euclidean signature.
Further, we choose to model the natural ultraviolet (UV) cutoff that is widely expected at the Planck scale \cite{Rovelli:1997qj,Witten_EveryPhysicist,Carlip_BriefHistory, Loll:2022xp,Garay_MinimumLength,Hossenfelder_MinimumLength} as a cutoff, 
$\bar{\Lambda}$, on the spectrum of the Laplace operator $\Delta$. This model of a natural UV cutoff has the advantage that it is simple, that it can also be implemented with the Lorentzian signature, see, e.g., \cite{chatwin2017natural,chatwindavies-kempf-simidzija2023,chatwindavies2023predictions}) 
and that it possesses an information-theoretic interpretation as a covariant bandlimitation \cite{kempf2000fields,kempf2004covariant,kempf2008information,kempf2009information,aasen2013shape}: it allows one to describe spacetime as simultaneously discrete and continuous \cite{kempf2010spacetime} in mathematically the same way that bandlimited information is simultaneously discrete and continuous. To see this, recall Shannon's sampling theorem which is in ubiquitous use throughout signal processing, \cite{shannon1948mathematical,cover1991information}.
{This theorem} holds that if the amplitudes of a bandlimited function 
are known on an arbitrarily chosen, sufficiently \footnote{A lattice is sufficiently dense if the average sample spacing is at most $1/(2\Omega)$ where $\Omega$ is the bandlimit.}  
dense lattice, then the function's amplitudes can be \it perfectly \rm  reconstructed everywhere. Further, this choice of UV cutoff renders the dimension, $N$, of the Hilbert space of scalar fields on the spacetime finite. 
By a spectral geometric result of Gilkey and Hawking \cite{Gilkey, Hawking}, $N$ can then be expressed in terms of a curvature expansion which is proportional to the Einstein-Hilbert action with higher order curvature corrections:
\begin{equation}
N = \frac{1}{16\pi^2}\int d^4x \sqrt{g}\left(\frac{\bar{\Lambda}^2}{2}+\frac{\bar{\Lambda}}{6}R+O(R^2)\right)
\label{dimH}
\end{equation}
Notice that this general form of \cref{dimH} had to be expected since $N$, being a curvature-dependent geometric invariant, should possess an expansion in scalar integrals over curvature scalars \cite{mukhanov2007introduction}. 
\cref{dimH} allows us to express the gravitational action basis independently as a trace, 
\begin{equation}\label{Einstein-Hilbert}
    S_g=\mu N= \mu \Tr(\mathbb{1}),
\end{equation} 
with $\mu= \frac{6 \pi}{\bar{\Lambda}}$. 
Since the higher order curvature corrections are suppressed by powers of $\bar{\Lambda}$, they are  unobservably small at currently experimentally accessible energies. What speaks in favor of including these terms
is that they render the gravity action $S_g$ of \cref{Einstein-Hilbert} not only very simple but also positive, since $N\in $ I\!N. 
This avoids the problem of the potential lower unboundedness of the Euclidean gravity action. We recall also that this type of correction terms is in any case necessarily induced through the renormalization of interacting fields \cite{Sakharov1, Sakharov2}. 

We now also write the action of $N_b$ free bosonic field species basis independently:
\begin{equation}
S_{b} = \frac{1}{2}\sum_{i=1}^{N_b} \mathrm{Tr}\left((\Delta + m^2) | \phi \rangle_i \langle \phi |_i \right)
\label{ScalarAction}
\end{equation}
For simplicity, we initially choose all species to possess the same nonzero mass, $m$.  
The bra-ket notation is here used to basis-independently denote fields in the Hilbert space of fields that we will sum over in the path integral.
The conventional representation of $S_b$ in coordinates would be obtained by performing the trace in a position basis. Here, we will not make the assumption that position bases exist. Instead, we perform the trace in the eigenbasis of $\Delta$, which is guaranteed to exist by the spectral theorem, to obtain:
\begin{equation} \label{sb}
    S_{b} = \sum_{i=1}^{N_b}\sum_{n=1}^N\lambda_n(\phi^{i}_n)^2.
\end{equation}
The $\{ \lambda_n \}, \ n \in \{1,..., N\}$ are the eigenvalues of the wave operator $\Delta+m^2$, which are positive, and the $\phi^{i}_n$ are the coefficients of $\vert \phi^{i}\rangle$ in the eigenbasis basis of $\Delta$. 
Similarly, the Dirac action of fermionic fields in a positional basis
\begin{equation} 
    S_f=\int d^4 x \sqrt{g} \bar{\Psi} \left(i \Gamma^{\mu} D_{\mu} \right) \Psi.
\end{equation}
we here express in the eigenbasis of the Dirac operator, reading, for $N_f$ fermionic species:
\begin{equation}
    S_f=\sum_{i=1}^{N_f}\sum_{n=1}^N \sqrt{\lambda_i} \theta^i_n\bar{\theta}^i_n
    \label{sf}
\end{equation}
For now, the fermionic wave operator in \cref{sf} is for simplicity taken to be the positive square root of the bosonic wave operator. 
The $\theta_n^i$, $\bar{\theta}^i_n$ are the Grassmann-valued components of the Dirac field $\psi^i$ of the $i$'th fermionic species.
In the eigenbasis of the wave operators, the total action, $S$, of gravity and matter therefore reads 
\begin{equation}
    S=S_g+S_b+S_f \label{totalaction}
\end{equation}
with $S_g$, $S_b$, and $S_f$ given by \cref{Einstein-Hilbert,sb,sf}.\newline 
\bf Rationale for working in the eigenbasis of the wave operators. \rm  
We choose to work in the eigenbasis of the wave operators for two reasons. First, in this basis, the action is expressed entirely in terms of diffeomorphism invariant quantities. Therefore, we avoid the problem, \cite{Kiefer, Hawking}, of path integrating over the metric while modding out the diffeomorphism group.
 
Second, when writing the action basis independently, the existence of a coordinate basis is no longer enforced, while 
wave operators, being self-adjoint, always possess an eigenbasis. Working in this basis, we will be able to show below that, besides regular regimes that describe a spacetime and matter, the path integral also contains physical regimes that do not admit position bases. These regimes are pre-geometric in the sense that they do not admit a mathematical representation of the degrees of freedom in terms of fields that live on a spacetime.  Consistent with the fact that the path integral of the action of \cref{totalaction} describes both geometric and pre-geometric regimes is that its invariance group is now manifestly enlarged from the diffeomorphism group to the full unitary group of the Hilbert
space of fields, since $S$ in \cref{totalaction} is basis independent.
(Recall that the unitary group is larger because, while coordinate transformations are unitary, not all unitaries are coordinate transformations). 

\bf The path integral. \rm 
The gravitational part of the path integral is now the integral over the eigenvalues, $\{\lambda_i\}_{i=1}^N$, of the wave operators,
and the sum over the number, $N$, of these eigenvalues. The matter part of the path integral is the integration over the coefficients $\phi_i^j$, $\theta_i^j,\bar{\theta}_i^j$ of the bosonic and fermionic fields.  

A choice to be made is whether to path integrate over spacetimes with or without boundaries. The choice to integrate effectively only over spacetimes with boundary can be implemented by integrating all eigenvalues of $\Delta$ from zero to the cutoff. The probability that an eigenvalue is zero is then of measure zero. Since zero is an eigenvalue of all compact Riemannian manifolds without a boundary (the corresponding eigenfunctions being constant functions), this effectively excludes these manifolds. 
Alternatively, and this will be the choice here, we can choose to enforce that zero is an eigenvalue of $\Delta$
by setting the lowest eigenvalue of the bosonic wave operator to $\lambda_1=m^2$. The calculations for spacetimes with boundaries are analogous. We therefore set $\lambda_1=m^2$ and
integrate over the Laplacian's remaining $(N-1)$ eigenvalues $\{\lambda_n\}, n=2,...,N$. Defining $\Lambda:=\bar{\Lambda}+m^2$, the path integral $Z$ then reads:
\begin{equation}
    Z=\sum^\infty_{N=1}\int_{m^2}^\Lambda \mathcal{D}\lambda \int \mathcal{D} \phi \int \mathcal{D}\theta \mathcal{D}\bar{\theta} e^{-\beta S} \frac{\Lambda^{N(\frac{N_f}{2}-1)}}{(N-1)!}.
    \label{PartFct}
\end{equation}
The powers of the UV cutoff $\Lambda$ arise because $Z$ is unitless. They could be omitted by choosing natural units in which $\Lambda=1$.
In \cref{PartFct}, we integrate over the $\lambda_i$ without ordering the eigenvalues, which we then remedy with the factor $((N-1)!)^{-1}$ to prevent overcounting the spectra. 
After integrating out the fermion and boson fields, as well as the Laplacian's spectrum $\{\lambda_n\}$ and after summing over $N$, the path integral evaluates to:
\begin{equation}
Z=C\ m^{d-2}\mathrm{exp} \left[ 2\ C\ \frac{\Lambda^{d/2}-m^{d}}{d} \right]
\label{UnorderedPartFct}
\end{equation}
Here, we defined $d := 2 - N_b +N_f$ and $    \beta_{max} := \frac{2N_f-N_b}{2\mu}$, as well as:
$    C := (2\pi)^{\frac{N_b}{2}} e^{-\beta \mu} \beta^{\mu \beta_{max}}\Lambda^{1-\frac{N_f}{2}}$. 
We can now use $Z$ to calculate expectation values.

\bf Spacetime dimension depends on the balance of bosonic and fermionic matter species. \rm We begin by showing that the effective dimension, $d_{eff}$, of the spacetime, whenever a spacetime representation does exist, depends on the numbers of fermion and boson species:
\begin{equation}\label{DimEff}
d_{eff}=d=N_f-N_b+2.
\end{equation}
To see this, we use Weyl's asymptotic formula. Weyl showed  that in regimes in which the behavior of the eigenfunctions of $\Delta$ is dominated by the spacetime dimension rather than by curvature, such as for large $\lambda$, the scaling law for the density $\rho(\lambda)$ of eigenvalues is that of flat space and is, therefore, in one-to-one correspondence to the dimension, $n$, of the manifold \cite{Kac,datchev2011inverse}:  
\begin{equation}
\lim_{\lambda \rightarrow \infty} \ \ \rho(\lambda) \propto \lambda^{n/2-1}.
    \label{ScalingRho}
\end{equation}
Here, we can calculate the scaling of $\rho(\lambda)$ from the eigenvalue probability density $p(\lambda_i)$ through
\begin{equation}
    \rho(\lambda) \propto p(\lambda_i)=\frac{1}{Z}\sum_N \int \mathcal{D}' \lambda \int \mathcal{D}\phi\int \mathcal{D}\theta \int \mathcal{D}\bar{\theta} e^{-\beta S},
    \label{FBProb}
\end{equation}
where $\mathcal{D}' \lambda$ includes all eigenvalues except one arbitrary $\lambda_i$, which we will call $\lambda$.  Evaluating these integrals and the sum yields: 
\begin{equation}
     p(\lambda) \propto \lambda^{N_f/2-N_b/2}.
\end{equation}
Comparing this scaling with  \cref{ScalingRho}, we  find that the effective dimension $d_{eff}$ is given by the number of fermionic and bosonic species in the model, as stated in Eq. \eqref{DimEff}. Notice that \cref{DimEff} could be interpreted as tracing the integer nature of spacetime dimensions to the integer nature of the numbers of species.

\bf The balance of $N_f$ and $N_b$, and therefore the spacetime dimension, can depend on the energy scale. \rm
While the calculations can be performed for positive and negative $d_{eff}$, a representation in terms of matter fields that live on a curved spacetime can of course only exist, i.e., a spacetime populated by matter can only emerge, if $d_{eff}>0$, i.e., if in \cref{DimEff}, $N_f>N_b-2$. 
More generally, we now show that, if the bosonic and fermionic species possess nontrivially distributed rest masses, then the numbers of bosonic and fermionic fields that effectively contribute to $N_f$ and $N_b$ are energy scale dependent. This in turn makes the effective dimension, $d_{eff}$, energy scale dependent.
To see this, we calculate the probability density $p(\lambda)$ in the case of $N_b$ families of bosons and $N_f$ families of fermions with different masses, to obtain:
\begin{equation}
    p(\lambda) \propto \prod^{N_b}_k(\lambda+m^2_{b_k})^{-1/2}\prod^{N_f}_l(\sqrt{\lambda}+m_{f_k}).
\end{equation}
In any interval of the energy axis, i.e., in any region of $\lambda$ values, in which the scaling of $p(\lambda)$ is approximately constant, \cref{DimEff} then yields an effective dimension $d_{eff}(\lambda)$:
\begin{equation}
    d_{eff}(\lambda)=-2\lambda\frac{\partial \log(p(\lambda))}{\partial\lambda}+2
\end{equation}
In energy regimes, i.e., in ranges of $\lambda$, in which $d_{eff}(\lambda)$ is not close to an integer, one of two cases can occur: (a) Weyls' formula does not apply because curvature at that scale is still large enough to prevent the eigenvalues to scale as in a flat space of some dimension, or (b) Weyl's formula does not apply because there is no spacetime representation.  

\begin{figure}[h!]
\begin{center}
\includegraphics[width = \linewidth]{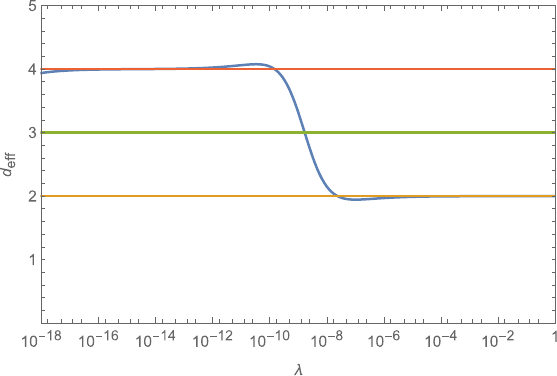}
\end{center}
\caption{ The effective dimension $d_{eff}(\lambda)$ for an example with $N_f=N_b=33$ where 32 fermionic masses and 30 bosonic masses are chosen to be of the order of $\lambda \sim 10^{-22}$ and 3 bosonic masses and one fermionic mass are chosen of the order of $\lambda \sim 10^{-10}$, in natural units where $\Lambda=1$.}
\label{DimRed}
\end{figure}

In \cref{DimRed}, we give an example with a set of species possessing low masses and a set of species with medium masses. At very low energies, $\lambda$, we expect as usual curvature-induced deviations from Weyl scaling, i.e., a case (a). Indeed, on the very left of the plot, we have $d_{eff}$ initially slightly below an integer, here $4$. 
For larger $\lambda$, one reaches Weyl scaling with $d_{eff}=4$. Further, when approaching the energy scale of the set of species with medium mass, we expect the balance of the number of fermionic and bosonic species to shift.
\cref{DimRed} indeed shows a transitional regime where, consistent with a case (b),
there exists no spacetime representation while $d_{eff}$ drops to $2$. Finally, toward the UV, one expects Weyl scaling to set in again and \cref{DimRed} indeed shows that one arrives at a spacetime of dimension $2$. While the case of dimensional reduction at high energies has been motivated in many models of quantum gravity \cite{Carlip_2017}, both increases and drops of $d_{eff}$ can here be modelled with suitable boson and fermion mass spectra. 
Notice also that just before the transition of  $d_{eff}$ from $2$ to $4$, $d_{eff}$ slightly overshoots $4$. It was pointed out in \cite{Calcagni2,Calcagni3}, that such an overshoot could potentially cause an observable effect in gravitational wave detectors. Finally, note that if we chose instead $N_f$ and $N_b$ with $N_f+2\le N_b$, then we would obtain a pre-geometric regime in the UV, with $d_{eff}<0$.   

\bf Expected number $\langle N\rangle$ of degrees of freedom per species. \rm
We now calculate the expectation value $\langle N \rangle$ of the number of eigenvalues, i.e., the expected dimension of the Hilbert space of fields for each species. $\langle N\rangle$ is of interest because it represents the expected number of `degrees of freedom' of each field in the given spacetime. To see this, recall that functions, $f$, in an $N$-dimensional function space ${\cal H}$ possess $N$ degrees of freedom in the sense that to fully determine such a function $f$, it suffices to know the function's amplitudes $a_n=f(x_n)$ at $N$ generic points $x_n$. This is because these $N$ equations can be used to determine the $N$ coefficients of $f$ in a basis. 

\begin{figure}[h!] 
\begin{center}
\includegraphics[width = \linewidth]{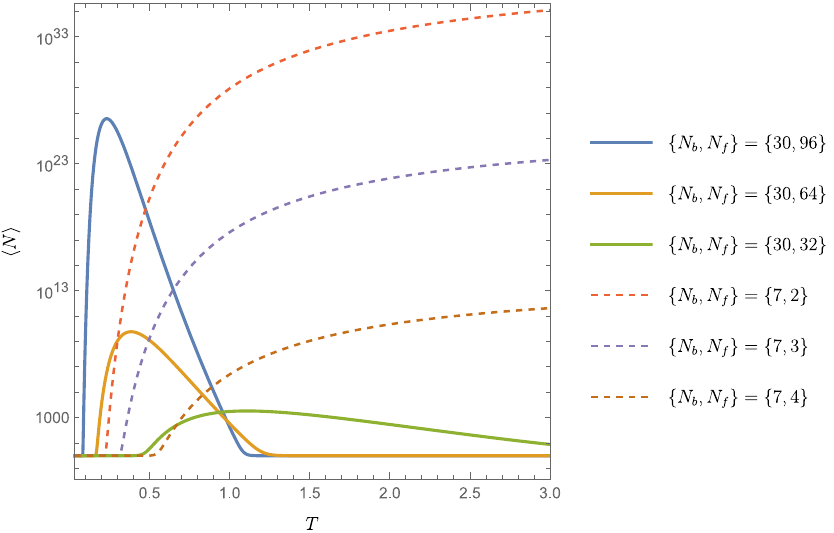}
\end{center}
\caption{Log plot of the expectation value $\langle N \rangle$ of the number of eigenvalues of $\Delta$ as a function of $T=1/\beta$, with $T$ displayed as a multiple of the UV cutoff $\Lambda$. Solid lines represent cases where $d_{eff}=N_f-N_b+2>0$, including the case $\{N_b,N_f\}={30,32}$ implying $d_{eff}=4$. In contrast, dashed lines represent cases where $d_{eff}<0$, i.e., where no spacetime representation exists and $\langle N\rangle$ grows without bound.}
\label{ExN}
\end{figure}

Since the calculation of $\langle N\rangle$ with nontrivial mass spectra is lengthy, we here for simplicity return to the case of one regime, where $d_{eff}$ is effectively constant, by setting all masses to a low energy value $m$. We obtain:
\begin{equation} \label{expN}
    \langle N \rangle = \frac{-Z^{-1}}{\beta}\frac{\partial Z}{\partial \mu} = 1 + 2 C~ \frac{\Lambda^{d/2}-m^{d}}{d}.
\end{equation}
Let us now analyze the dependence of $\langle N\rangle$ on $m,N_f,N_b$ and $\beta$, the latter playing the role of an inverse temperature $\beta=1/T$, if we choose to interpret the Euclidean path integral as a thermal partition function. $\langle N\rangle$ depends on $\beta$ through $C$, as defined with \cref{UnorderedPartFct}. 
From there we obtain that if the spacetime dimension is positive, $d_{eff}>0$, and $N_f>1$, then $\langle N \rangle \rightarrow 1$ for $\beta \rightarrow \infty$ and $\beta \rightarrow 0$ with $\langle N \rangle$ maximal at 
  $\beta=  \beta_{max} = (2N_f-N_b)/2\mu$.
From \cref{expN,UnorderedPartFct}, the mass $m$ influences only the height of the curve $\langle N\rangle(T)$. The shape of the curve, as shown in \cref{ExN}, is determined by the balance of $N_f$ and $N_b$, through $C$. 
\cref{ExN} shows that, 
for boson dominance in the sense that $d_{eff}=N_f-N_b+2\le 0$, i.e., when where there is no spacetime representation, the expected number of degrees of freedom $\langle N\rangle$ is unbounded while for fermion dominance, $d_{eff}=N_f-N_b+2> 0$, i.e., when a spacetime representation exists, the expected number of degrees of freedom $\langle N\rangle$ is bounded. This result is intuitive since when there exists a representation on a spacetime and the spacetime has a finite volume, then fermions should indeed only allow a finite number of degrees of freedom in that volume, given the UV cutoff. Let us now investigate the expected spacetime volume.

\begin{figure}
\begin{center}
\includegraphics[width = \linewidth]{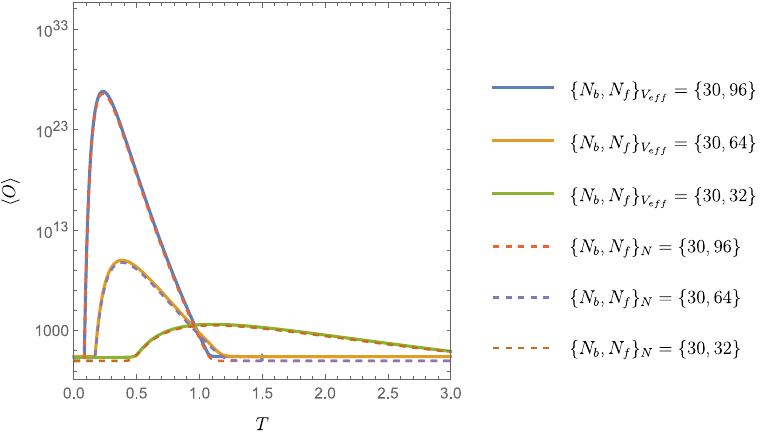}
\end{center}
\caption{Log plot of a comparison between the expectation value of the Hilbert space dimension $\langle N \rangle$ (dashed lines) and the effective  $V_{eff}$ (solid lines) in natural units, $\Lambda=1$.  $N_f$ and $N_b$ are chosen as in the three cases of fermion domination in \cref{ExN}. We see that the effective volume and effective Hilbert space dimension are of the same order of magnitude, providing a quantum generalization of \cref{dimH}}
\label{ExL}
\end{figure}

\bf The spectral gap and the size of the spacetime. \rm For a compact Riemannian manifold $\mathcal{M}$, the spectral gap of $\Delta$ is closely related to $\ell^{-2}$, with $\ell$ its largest geodesic distance (see e.g. \cite{HE}) and therefore, roughly, to its volume $V\approx\ell^d$. 
In this sense, defining $g:=\lambda_2-\lambda_1=\lambda_2-m^2$, the expression $\langle g \rangle^{-d/2}$ provides a rough estimate of the effective volume $V_{eff}\equiv\langle g \rangle^{-d/2}$ of 
an emergent spacetime of dimension $d$.
To this end, we calculate:
\begin{equation}
    \langle \lambda_2 \rangle = \int_m^\Lambda d \lambda_2 \ \lambda_2 P(\lambda_2|N\geq2).
\end{equation}
Here, $P(\lambda_2|N\geq2)$ is the probability of the value $\lambda_2$, assuming $N \ge 2$. This probability can be found using Bayes' theorem: $P(\lambda_2|N\geq2) = \frac{P(\lambda_2, N\geq2)}{P(N\geq2)} = \frac{P(\lambda_2, N\geq2)}{1-P(N=1)}$.
We find,
\begin{align}
    \langle \lambda_2 \rangle &= \frac{A(\frac{d}{2C})^{2/d} \Gamma(1+\frac{2}{d},C\frac{2m^{2}}{d})-\Gamma(1+\frac{2}{d},2C\frac{\Lambda^{d/2}}{d}))}{C},\notag\\
     & \mbox{ where~~~ } A = C\frac{\exp{2C \frac{m^{d}}{d}}}{1 - 
 \exp{-2C \frac{\Lambda^{d/2} - m^{d}}{d}}}.
\end{align}
Here, $\Gamma(a,b)=\int_b^\infty dy \ y^{a-1}e^{-y}$ is the incomplete gamma function. In \cref{ExL}, we plot the resulting effective spacetime volume $V_{eff}$ as function of $T$ in natural units, along with $\langle N \rangle$. Analytically, the curves of 
$\langle N\rangle$ and $V_{eff}$ are not identical, but they match very closely. The ratio $\langle N \rangle$/$V_{eff}$ varies but stays roughly of order 1 and the maximum of $\langle N \rangle$ and $V_{eff}$ is at the same value, $\beta_{max}$.



\bf Consistency check. \rm
 \rm 
The fact that the curves for the number of degrees of freedom and the volume closely match, see \cref{ExL},  implies a constant density of degrees of freedom, up to corrections, which constitutes a highly nontrivial consistency check. To see this, notice that in \cref{dimH} the integral over the first term on the RHS is the spacetime volume. Dividing \cref{dimH} by the volume therefore shows that, for a classical spacetime, the density of degrees of freedom $N/V$ of a classical field is constant, up to curvature corrections: $N/V=\bar{\Lambda}^2/32\pi^2 + \int dV O(R)$. The fact that the present calculations yield an analogous result after quantization is nontrivial since here in the quantum case the density of degrees of freedom was instead indirectly inferred using Weyl's law and the behavior of the gap.
Indeed, in the non-geometric regime where  $N_b>2+N_f$, i.e., where there is no positive effective spacetime dimension, there is no geometric relationship between $V_{eff}$ and $\langle N \rangle$ since for large $T$, the quantity $V_{eff}$ remains finite while $\langle N \rangle$ diverges.
\newline
\bf Outlook. \rm 
Passing this nontrivial consistency check, the present results on the emergence of spacetimes, and possible changes of their dimension, support the proposal in \cite{Kempf1}. There, it was proposed that the physics of the emergence of spacetime and matter can be described, technically, as the emergence of the mathematical representability of otherwise abstract degrees of freedom as quantum fields on a curved spacetime. Further, the present results now encourage exploring the use of the Lorentzian signature and the inclusion of gauge symmetries and interactions.
As we will show in a longer follow-up paper, to include interactions, even if only perturbatively, is 
the key \cite{Kempf1,Kempf2016} to moving to the Lorentzian signature. 
This is because, in the Lorentzian signature, neither Weyl scaling nor the gap can be relied upon to infer the dimension or volume of an emergent spacetime. The use of interaction terms, however, will allow one to calculate even more, namely the metric itself.  
The reason is that interaction terms are local, i.e., they single out positional bases, whenever they exist, namely as those bases in which the interaction terms are diagonal. This then allows one to represent propagators in positional bases. At short spacetime distances, a propagator is a function of the geodesic distance, therefore also yields infinitesimal distances and, therefore, the metric \cite{Kempf1,Kempf2016}.



It will also be interesting to explore the non-geometric regimes. For example, the spectra of wave operators are naturally linked in geometric regimes but may vary independently in non-geometric regimes. Indications for this possibility have been found recently with methods of Causal Dynamical Triangulations, where it was shown that the scaling of the spectra of the Laplacians on $k$-forms can vary with $k$ \cite{Reitz}.
One may also explore path integrating over mass spectra and over $N_f$ and $N_b$.

\bf Acknowledgements: \rm AK acknowledges support through a Discovery Grant by the National Science and Engineering Research Council of Canada (NSERC), a Discovery Project Grant from the Australian Research Council (ARC) as well as two Google Faculty Research Awards. MR was supported in part by the Excellence Initiative - Research University Program at the Jagiellonian University in Krakow and the National Science Centre, Poland, under grant no. 2019/33/B/ST2/00589. B\v{S} is supported in part by the Perimeter Institute, which is supported in part by the Government of Canada through the Department of Innovation, Science
and Economic Development Canada and by the Province of Ontario through the Ministry of Economic Development, Job Creation and Trade.

\bibliographystyle{apsrev4-2}
\bibliography{references}

\end{document}